\documentclass{ws-rv9x6}
\usepackage{subfigure}     
\usepackage[square]{ws-rv-van}     
\usepackage{hyperref}
\usepackage{bm}

\newcommand{\be}{\begin{equation}}
\newcommand{\ee}{\end{equation}}
\newcommand{\ba}{\begin{eqnarray}}
\newcommand{\ea}{\end{eqnarray}}

\newcommand{\rhonuc}{\mbox{$\rho_\mathrm{nuc}$}}

\begin{document}

\chapter[Pairing and the Cooling of Neutron Stars]{Pairing and the Cooling of Neutron Stars}

\author[Dany Page]{Dany Page}

\address{Instituto de Astronom\'ia, Universidad Nacional Aut\'onoma de M\'exico, \\
Ciudad Universitaria, M\'exico, D.F. 04510, Mexico}

\begin{abstract}
In this review, I present a brief summary of the impact of nucleon pairing at supra-nuclear densities on the cooling of
neutron stars.
I also describe how the recent observation of the cooling of the neutron star in the supernova remnant Cassiopeia A 
may provide us with the first direct evidence for the occurrence of such pairing.
It also implies a size of the neutron $^3\! P\! - \! F_2$ energy gap of the order of 0.1 MeV.
\end{abstract}

\body

\section{Introduction}

With radii of the order of 10 km and masses between 1 and 2 $M_\odot$, neutron stars a giant nuclei the size of a city.
Neutron star matter at densities around \rhonuc, the nuclear saturation density, is certainly made of neutrons
and protons, with electron and muons to guarantee charge neutrality.
At higher densities it is likely that hyperons and/or deconfined quarks are present,
as well as charged meson condensates.
The presence of such ``exotica'' can have a dramatic effect on the equation of state (EOS),
softening it and reducing the maximum possible neutron star  mass.
The recent measurement of a two-solar-mass neutron star is, however, imposing very severe constraints on the EOS
\cite{Demorest:2010fk}.
The chemical composition of dense matter also has a strong effect on the cooling of a neutron star, by determining
the allowed neutrino emission channels, a fact that provides another handle on the EOS.
However, the occurrence of pairing, which strongly affects excitations but has only a very weak effect on the EOS,
dramatically alters the neutrino emissivity of the star and hence its cooling.
Comparison of theoretical models with observations almost certainly requires the occurrence of pairing,
and the recent observation of the rapid cooling of the neutron star in Cassiopeia A \cite{Heinke:2010zr}
may be the first direct evidence for it.

In this chapter, I will present a brief description of the physical processes relevant for the modeling of neutron star
cooling followed by a series of generic results, and an overview of the ``Minimal Cooling'' paradigm and 
its application to Cas A.
This presentation is necessarily short and the reader can find more details in the reviews 
\cite{Yakovlev:2004kx,Page:2006fk,Page:2009uq}.
An introduction to the physics of neutron stars can be found in \cite{Page:2006kx}.

\section{The Basic Physics of Neutron Star Cooling}

Most neutron star cooling calculations involve solving numerically the heat transport and energy balance
equations, in their fully general-relativistic form (see, e.g., Appendix B in \cite{Page:2004zr}). 
However, simple global energy-balance (in its Newtonian form to keep it simple) as
\be
\frac{dE_{th}}{dt} = C_V \frac{dT}{dt}
                   = -L_\nu - L_\gamma + H
\label{Eq:cooling}
\ee
illustrates the main features, where it is assumed that the star's interior is isothermal with temperature $T$,
a state reached within a few decades after its birth in a core-collapse supernova.
Here, $E_{th}$ is the star's total thermal energy, $C_V$ its specific heat,
and $L_\gamma$ and $L_\nu$ its photon and neutrino luminosities, respectively. 
The term $H$, for ``heating'', represents possible dissipative processes which will not be considered here.

\subsection{Specific heat}

The core provides most of the specific heat and since it consists of degenerate fermions one naturally has
$C_V \propto T$.
When matter is made only of nucleons and leptons, about 70\% of $C_V$ is provided by the neutrons,
20\% by the protons and 10\% by the leptons.
For simple numerical estimates one can use $C_V \simeq 10^{39} \,T_9$ erg K$^{-1}$,
where $T_9 \equiv T/10^9$ K.
Notice, however, that in the presence of pairing $C_V$ can be strongly suppressed.

\subsection{Photon emission and the envelope}

The surface photon luminosity is traditionally written as
\be
L_\gamma = 4 \pi R^2 \, \sigma_{\rm SB} T_e^4
\label{Eq:Lphot}
\ee
which {\em defines} the effective temperature $T_e$ ($R$ is the star's radius and $\sigma_{\rm SB}$ the
Stefan-Boltzmann constant).
$T_e$ is much lower than the star's interior temperature: even in the case the interior is isothermal
there always exits a strong temperature gradient in the uppermost layers, commonly called the {\em envelope}.
As a rule of thumb one can use
\be
T_e \approx 10^6 (T/10^8 {\rm K})^{1/2} \; {\rm K}
\label{Eq:Tb-Te}
\ee
which implies that $L_\gamma \approx 10^{35} \, T_9^2$ erg s$^{-1}$.
The details of this $T_e - T$ relationship depend on the chemical composition of the envelope
and the presence or absence of a strong magnetic field.

\subsection{Neutrino emission processes}

\begin{table} [t]
\tbl{Dominant neutrino emission processes.}
{\begin{tabular}{llcc} 
\hline 
     Name           &               Process              &       Emissivity$^\dagger$       &  Efficiency    \\ 
                    &                                        &   (erg cm$^{-3}$ s$^{-1}$) &               \\
\hline 
\parbox[c]{3.0cm}{Modified Urca cycle\\ (neutron branch)} &
\rule[-0.25cm]{0.02cm}{0.60cm}
$\begin{array}{l} n+n \rightarrow n+p+e^-+\bar\nu_e \\ n+p+e^- \rightarrow n+n+\nu_e \end{array}$  & 
$\sim 2\!\! \times \!\! 10^{21} \: R \: T_9^8$ & Slow \\
\parbox[c]{3.0cm}{Modified Urca cycle\\ (proton branch)}  &
\rule[-0.25cm]{0.02cm}{0.60cm}
$\begin{array}{l} p+n \rightarrow p+p+e^-+\bar\nu_e \\ p+p+e^- \rightarrow p+n+\nu_e \end{array}$  & 
$\sim 10^{21} \: R \: T_9^8$ & Slow \\
Bremsstrahlung          &
$\begin{array}{l} n+n \rightarrow n+n+\nu+\bar\nu \\ n+p \rightarrow n+p+\nu+\bar\nu \vspace{-0.0cm} \\
    p+p \rightarrow p+p+\nu+\bar\nu \end{array}$                                     & 
$\sim 10^{19} \: R \: T_9^8$  & Slow \\
\parbox[c]{3.0cm}{Cooper pair \\ formations}          &
$\begin{array}{l}    n+n \rightarrow [nn] +\nu+\bar\nu \\ p+p \rightarrow [pp] +\nu+\bar\nu \end{array}$  & 
$\begin{array}{l} \sim 5\!\!\times\!\! 10^{21} \: R \: T_9^7 \\ \sim 5\!\!\times\!\! 10^{19} \: R \: T_9^7 \end{array}$ & Medium \\
\parbox[c]{3.0cm}{Direct Urca cycle\\ (nucleons)} &
\rule[-0.25cm]{0.02cm}{0.60cm}
$\begin{array}{l} n \rightarrow p+e^-+\bar\nu_e \\ p+e^- \rightarrow n+\nu_e \end{array}$              & 
$\sim 10^{27} \: R \: T_9^6$ & Fast \\
\parbox[c]{3.0cm}{Direct Urca cycle\\ ($\Lambda$ hyperons)} &
\rule[-0.25cm]{0.02cm}{0.60cm}
$\begin{array}{l} \Lambda \rightarrow p+e^-+\bar\nu_e \\ p+e^- \rightarrow \Lambda+\nu_e \end{array}$              & 
$\sim 10^{27} \: R \: T_9^6$ & Fast \\
\parbox[c]{3.0cm}{Direct Urca cycle\\ ($\Sigma^-$ hyperons)} &
\rule[-0.25cm]{0.02cm}{0.60cm}
$\begin{array}{l} \Sigma^- \rightarrow n+e^-+\bar\nu_e \\ n+e^- \rightarrow \Sigma^-+\nu_e \end{array}$              & 
$\sim 10^{27} \: R \: T_9^6$ & Fast \\%
$\pi^-$ condensate &$n+<\pi^-> \rightarrow n+e^-+\bar\nu_e$  &
$\sim 10^{26} \: R \: T_9^6$  & Fast \\
$K^-$ condensate   &$n+<K^-> \rightarrow n+e^-+\bar\nu_e$  &
$\sim 10^{25} \: R \: T_9^6$  & Fast \\
\parbox[c]{3.0cm}{Direct Urca cycle\\ (u-d quarks)} &
\rule[-0.25cm]{0.02cm}{0.60cm}
$\begin{array}{l} d \rightarrow u+e^-+\bar\nu_e \\ u+e^- \rightarrow d+\nu_e \end{array}$              & 
$\sim 10^{27} \: R \: T_9^6$ & Fast \\
\parbox[c]{3.0cm}{Direct Urca cycle\\ (u-s quarks)} &
\rule[-0.25cm]{0.02cm}{0.60cm}
$\begin{array}{l} s \rightarrow u+e^-+\bar\nu_e \\ u+e^- \rightarrow s+\nu_e \end{array}$              & 
$\sim 10^{27} \: R \: T_9^6$ & Fast \\
\end{tabular}}
\begin{tabnote}
$^\dagger$ The coefficients $R$'s are control functions to incorporate the effects of pairing, see
\S~\ref{Sec:pairing_effects}.
\end{tabnote}
\label{Tab:nu}
\end{table} 

A list of the most important neutrino emission processes is presented in Table~\ref{Tab:nu},
with rough values of their emissivities.
They are separated into ``slow'' and ``fast'' processes, the former involving 5 and 
the latter only 3 degenerate fermions.
Notice the different temperature dependences: $T^6$ for the three degenerate fermion processes
compared to $T^8$ for the five fermion ones, a direct consequence of the stronger phase space limitation
resulting in a significantly reduced emissivity.
The Cooper pair process is described in \S~\ref{Sec:pairing_effects}.
A detailed description of neutrino processes can be found in \cite{Yakovlev:2001vn}
and an alternative approach in \cite{Voskresensky:2001ys}.

\subsection{Effects of nucleon pairing}
\label{Sec:pairing_effects}

The occurrence of pairing, either of neutron or of protons
\footnote{Hyperons, and deconfined quarks, if present,
are also expected to pair. I will, here, only consider nucleons but most of what follows naturally translates
to these others cases.}, 
introduces a series of important effects:
\begin{description}
\item[\rm A)] 
Alteration, and possible strong suppression when $T \ll T_c$, of the specific heat of the paired component.
\item[\rm B)] 
Reduction, and possible strong suppression when $T \ll T_c$, of the emissivity of neutrino processes 
the paired component is involved in.
\item[\rm C)] 
Triggering of the ``Cooper pair breaking and formation'' (PBF) neutrino process which is  
very efficient in the case of spin-triplet pairing.
\end{description}
These effects are direct consequences of the alteration of the quasi-particle spectrum
by the development of an energy gap $D(\bm k)$,
\begin{flalign} 
& \mathrm{normal \; phase:} \hspace{10mm} \epsilon(k) = v_F  (k-k_F)  &
\label{Eq:N_spectrum}
\\
& \mathrm{paired \; phase:} \hspace{11mm} \epsilon(\bm k) = \pm \sqrt{[v_F (k-k_F)]^2 + D^2(\bm k_F)} & 
\label{Eq:P_spectrum}
\end{flalign}
which severely limits the available phase space when $T \ll T_c$.
In cooling calculations these effects are introduced through ``control functions'':
\be
c_V \longrightarrow R_c \, c_V
\;\;\;\; \text{and} \;\;\;\;
\epsilon_\nu^X \longrightarrow R_X \, \epsilon_\nu^X \; .
\ee
There is a large family of such control functions for the various types of pairing and the numerous neutrino processes ``X''
(they are the factors $R$ in Table~\ref{Tab:nu}).
For nodeless gaps the $R$'s are Boltzmann-like factors $\sim \exp[-2\Delta(T)/k_B T]$ and result in a
strong suppression when $T \ll T_c$, while for gaps with nodes the suppression is much milder.
Regarding the specific heat, there is a sudden increase, by a factor $\sim 2.4$ at $T=T_c$, followed by a reduction
at lower $T$.

The effect C), neutrino emission from the formation and breaking of Cooper pairs \cite{Flowers:1976ly,Voskresenskii:1986qf},
can be seen as an inter-band transition (as, e.g., $n \rightarrow n + \nu \overline\nu$)
where a neutron/proton quasiparticle from the upper ($+$) branch of the spectrum of Eq.~(\ref{Eq:P_spectrum})
falls into a hole in the lower ($-$) branch
Such a reaction is kinematically forbidden by the excitation spectrum of the normal phase, Eq.~(\ref{Eq:N_spectrum}), 
but becomes possible in the presence of an energy-gap, Eq.~(\ref{Eq:P_spectrum}).
The resulting emissivity can be significantly larger than the one of the modified Urca process in the case of
spin-triplet pairing.

\subsection{Theoretical predictions of nucleon gaps}

Pairing is usually assumed to occur in a single spin-angular momentum channel $\lambda=(s,j)$ and
the gap function is a $2\times2$ matrix in spin space $\hat\Delta_\lambda$.
At low $k_{Fn}$, or $k_{Fp}$, it is theoretically predicted that the preferred channel is $\lambda=(0,0)$ in $S$-wave,
i.e., the spin-singlet $^1\! S_0$.
At larger Fermi momenta the $^1\! S_0$ interaction becomes repulsive and the preferred channel is $\lambda=(1,2)$
in $P$ and $F$ waves (the mixing being due to the tensor interaction), i.e., the spin-triplet $^3\! P\! - \! F_2$.
In the $^1\! S_0$ channel, which has also been called the ``A'' phase, the gap is spherically symmetric and 
$\hat\Delta_{(0,0)}$ depends on one single scalar $\Delta(k)$ so that the energy gap $D(k_F)$, at the Fermi surface,
is simply
\begin{flalign} 
& \mathrm{A \; phase} \; (^1\! S_0): \hspace{17mm}D(k_F) = \Delta(k_F) \; . & 
\end{flalign}
In the $^3\! P\! - \! F_2$ channel, $\hat\Delta_{\lambda}$
has contributions from all possible orbital angular momentum  $l$ and $m_j$ components, i.e.,
$\hat\Delta_\lambda = \sum_{l,m_j} \Delta^{m_j}_{l \, \lambda}(k)  \hat{G}^{m_j}_{l \, \lambda}(\hat{\bm k}_F)$
where the $\hat{G}^{m_j}_{l \, \lambda}(\hat{\bm k}_F)$ are $2 \times 2$ spin matrices describing
the angular dependence of $\hat\Delta$ which is thus {\em not} spherically symmetric. 
Microscopic calculations restricted to the $^3\! P_2$ channel indicate that the largest component of 
$\hat\Delta_{\lambda}$ corresponds to the $m_j =  0$ sub-channel or, possibly, the $m_j = \pm 2$ one,
sometimes called the ``B'' and ``C'' phases, respectively. 
For these two special cases the energy gap $D(\bm k_F)$ is given by  \cite{Amundsen:1985nx}
\begin{flalign} 
& \mathrm{B \; phase} \; (^3\! P_2, m_j=0): \hspace{5mm}
D^2(\bm k_F) = \frac{1}{2} \left[ \Delta^0_{2 \, \lambda}(k_F) \right]^2 \, \frac{1+3 \cos^2 \theta}{8\pi} & 
\\
& \mathrm{C \; phase} \; (^3\! P_2, m_j=\pm 2): \hspace{2mm}
D^2(\bm k_F) = \;\;\; \left[ \Delta^2_{2 \, \lambda}(k_F) \right]^2 \;\;\; \frac{3 \sin^2 \theta}{8\pi} & 
\end{flalign}
where $\theta$ is the angle between $\bm k_F$ and the arbitrary quantization axis.
The relationship between the phase transition critical temperature $T_c$ and the energy gap $D(\bm k_F)$
is approximately given by the usual result
\be
k_B T_c \approx  0.57 \; \overline\Delta(k_F;T=0)
\ee
for all three phases \cite{Amundsen:1985nx,Baldo:1992oq}, 
where $\overline\Delta(k_F ; T=0)$ is obtained by angle averaging of $D^2(\bm k_F)$ over the Fermi surface
\be
\left[ \, \overline\Delta(k_F;T=0) \right]^2 \equiv  \int \!\!\! \int \! \frac{d\Omega}{4\pi} \, D^2(\bm k_F;T=0) \; .
\ee
%

\begin{figure}
\centerline{\psfig{file=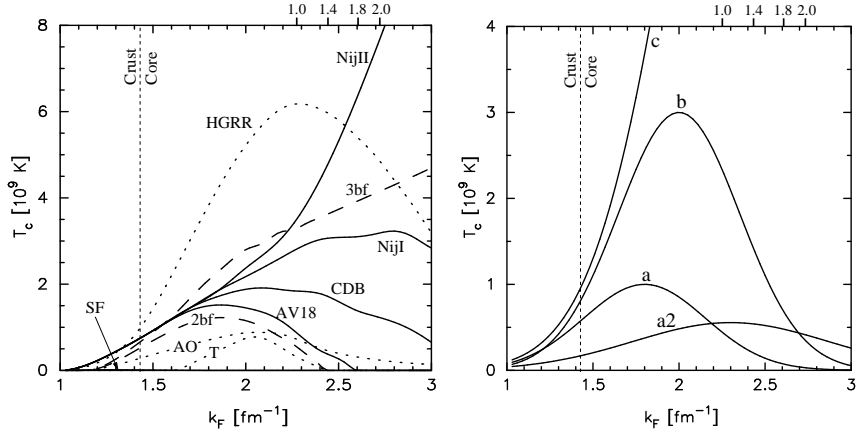,width=0.99\textwidth}}
\caption{Left panel: some theoretical predictions of $T_c$ for the neutron $^3\! P\! - \! F_2$ gap.
              See text for description.
              Right panel: some phenomenological models of $T_c$ for the neutron $^3\! P\! - \! F_2$ gap
              used in neutron star cooling simulations.
              Models ``a'', ``b'', and  ``c'' are from \cite{Page:2004zr} and \cite{Page:2009vn}, model  ``a2'' from \cite{Page:2011ys}.
              On the top margin are marked the values of $k_{Fn}$ at the center of a $1.0$, $1.4$, $1.8$, and $2.0$
              $M_\odot$ star built with the APR EOS \cite{Akmal:1998fk}.}
\label{Fig:Tc_n}
\end{figure}

\begin{figure}
\centerline{\psfig{file=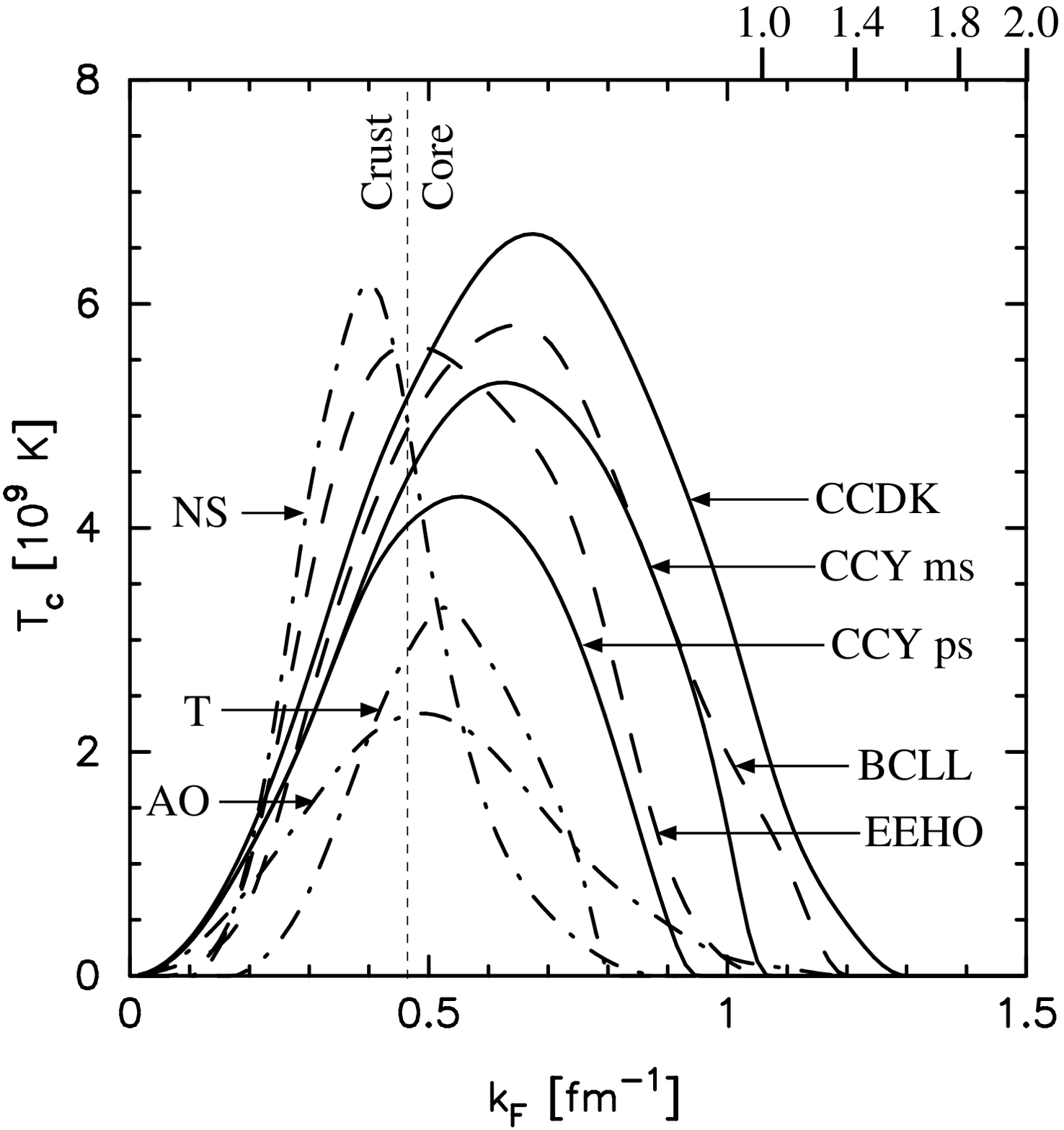,width=0.5\textwidth}}
\caption{Some theoretical predictions of $T_c$ for the proton $^1\! S_0$ gap. 
              See \cite{Page:2004zr} for references.
              On the top margin are marked the values of $k_{Fp}$ at the center of a $1.0$, $1.4$, $1.8$, and $2.0$
              $M_\odot$ star built with the APR EOS \cite{Akmal:1998fk}.}
\label{Fig:Tc_p}
\end{figure}

Figure~\ref{Fig:Tc_n} shows a sample of theoretical predictions of $T_c$ for the neutron $^3\! P\! - \! F_2$ gap.
The three dotted lines present some of the first published models:
``HGRR'' from \cite{Hoffberg:1970dq}, ``T'' from \cite{Takatsuka:1972cr} and ``AO'' from \cite{Amundsen:1985nx}.
The four continuous lines show models from \cite{Baldo:1998bh} and illustrate the uncertainty on the gap size due to the
problem that no $N\! -\! N$ potential
 reproduces the measured $^3\! P_2$ phase-shift above $E_\mathrm{lab} \simeq 300$ MeV
(translating into $k_{F n} \! \sim \! 1.8$ fm$^{-1}$).
These gaps were calculated with the Nijmegen II (``NijII''), Nijmegen I (``NijI''), CD-Bonn (``CDB''), and Argonne $V_{18}$
(``AV18'') potentials (displayed values are taken from the middle panel of Figure~4 of \cite{Baldo:1998bh}).
These calculations are at the ``BCS approximation'' level, i.e., do not include medium polarization effects.
In the case of the $^1\! S_0$ gap, medium polarization is known to result in screening and to reduce the size of the gap.
In the case of a $^3\! P_2$ gap, polarization with central forces is expected to result in anti-screening and to increase 
the size of the gap.
However, Schwenk \& Friman \cite{Schwenk:2004qf} showed that spin-dependent non-central forces
 do the opposite and strongly screen the coupling in the $^3\! P_2$ channel, resulting in a $T_c$ lower than
$10^7$ K: this ``SF'' value is indicated in the Figure by an arrow !
Finally, 3-body forces are know to be essential for both nuclear structure and neutron star matter.
They are repulsive in the bulk but at the Fermi surface in the $^3\! P\! - \! F_2$ channel they turn out to be strongly attractive.
The two dashed lines in Figure~\ref{Fig:Tc_n} present results from \cite{Zuo:2008ij} where the ``2bf'' model only considers 2-body
forces (from the Argonne $V_{18}$) while the ``3bf'' includes a meson exchange model 3-body force:
the result is a growing $^3\! P\! - \! F_2$ gap which shows no tendency to saturate at high density.

A set of theoretical predictions of $T_c$ for the proton $^1 \! S_0$ gap is shown in Figure~\ref{Fig:Tc_p}. 
Variations in the size of this gap are not as large as for the neutron triplet gap but uncertainty on the $k_F$ range 
is significant, translating in an uncertainty of a factor $\sim 4$ on the density range covered by proton superconductivity.

Given these large uncertainties on the size of the neutron $^3\! P\! - \! F_2$ gap (about three orders of magnitude)
and the fact that neutrino suppression depends on it through an exponential Boltzmann-like factor, this gap
is often considered as a free parameter in neutron star cooling models.
The extreme sensitivity of the cooling history on the size of this gap can be taken to one's advantage, by inverting the problem,
as it may allows us to {\em measure} it by fitting theoretical models to observational data \cite{Page:1992bs}.
The right panel of Figure~\ref{Fig:Tc_n} presents the phenomenological neutron $^3\! P\! - \! F_2$ gaps
used in the following cooling calculations.

\section{Some Illustrative Examples of Neutron Star Cooling}

\begin{figure}[t]
\centerline{\psfig{file=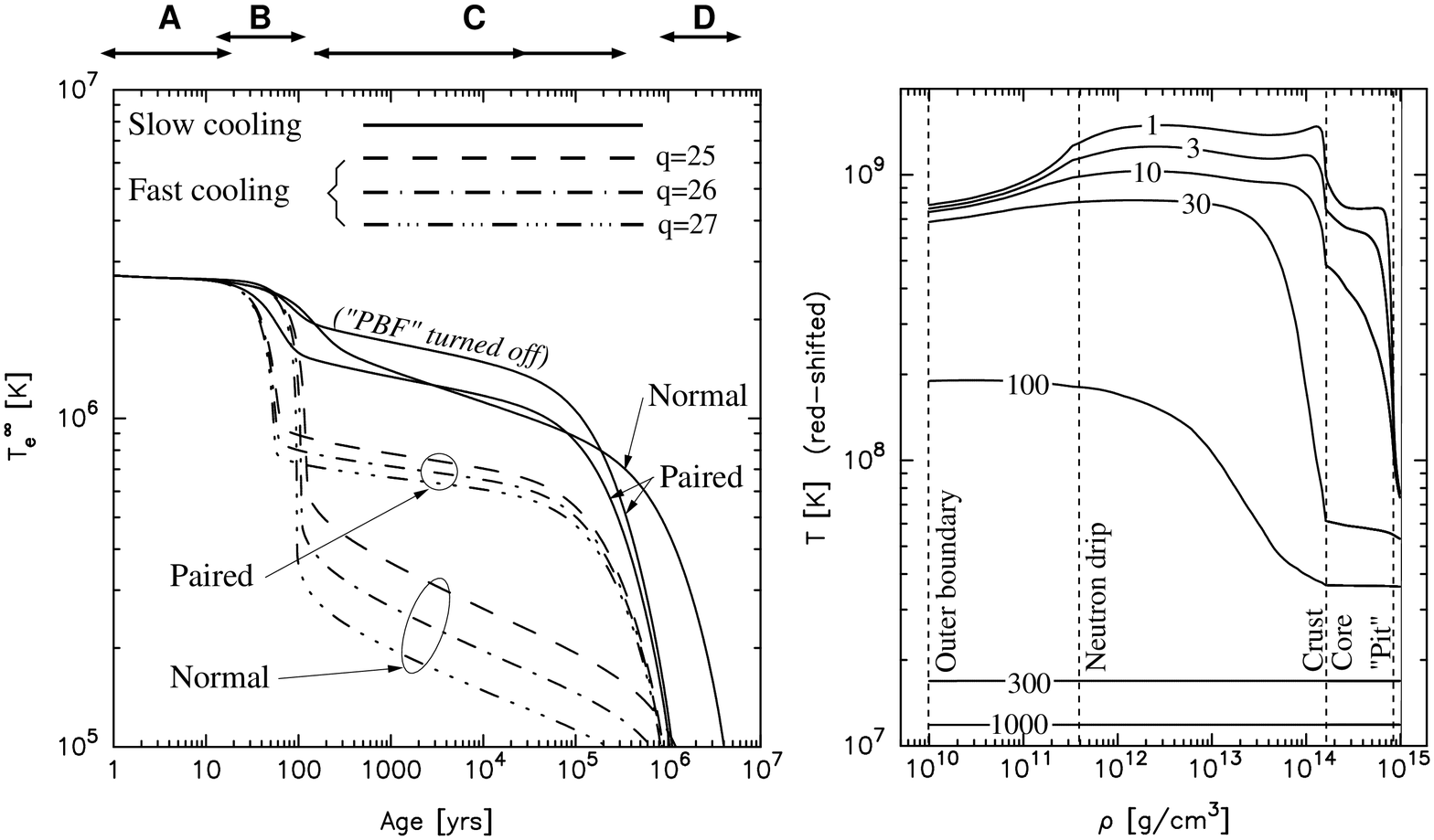,width=0.99\textwidth}}
\caption{Left panel: cooling curves for various illustrative cooling scenarios.
               Right panel: temperature profile evolution for the fast cooling model with $q=26$.
               The numbers on the curves give the age of the star, in years.
               See text for description.}
\label{Fig:Cool_Basic}
\end{figure}

To illustrate several of the possible behaviors of a cooling neutron star, I show in Figure~\ref{Fig:Cool_Basic}
a set of models, all based on the same star of mass $1.4 M_\odot$ built with the APR EOS \cite{Akmal:1998fk}.
The ``slow cooling'' models only include, in the core, the slow neutrino processes of Table~\ref{Tab:nu} as well as the
BPF process.
For the ``fast cooling'' models I added, at $\rho > 3 \rho_\mathrm{nuc}$, a fast process with emissivity 
$\epsilon_{Fast}^{(q)} = 10^q \cdot T_9^6$ erg cm$^{-3}$ s$^{-1}$, with $q=25$, $26$, and $27$,
that simulates the effect of a kaon condensate, a pion condensate, or a direct Urca, respectively.
These models, all based on the same EOS, are not self-consistent  but have the advantage that the only differences between them
is the presence/absence of the $\epsilon_{Fast}^{(q)}$ process and the presence/absence of pairing. 
The models with pairing include the neutron $^1\! S_0$ gap from \cite{Schwenk:2004qf} in the inner crust, a
$^1\! S_0$ proton gap in the outer core from \cite{Takatsuka:1973uq}, model ``T'' of Figure~\ref{Fig:Tc_p},
and the phenomenological neutron $^3\! P\! - \! F_2$ gap ``b'' of Figure~\ref{Fig:Tc_n}.

The various distinctive phases of the evolution are marked as ``A'', ``B'', ``C'', and ``D'' above the cooling curves.
Phases A and B are determined by the evolution of the curst while C and D reflect the evolution of core.
\\
{\bf Phase A:} the surface temperature $T_e$ is determined by the evolution of the outer crust only.
At such early stages the temperature profile in the outer crust is independent of what is happening deeper in the star and 
all models have the same $T_e$.
\\
{\bf Phase B:} the age of the star becomes comparable to the thermal relaxation time-scale of the crust and
heat flow controls the evolution of $T_e$. 
This thermal relaxation phase is depicted in the right panel of the Figure which shows the evolution of the $T$-profile
for the fast cooling model with $q=26$ in the absence of pairing (marked as ``Normal'' in the left panel).
One sees that very early on the ``pit'' (where the fast neutrino emission is occurring) is very cold and during the first
30 years heat is flowing from the outer core into the pit whose temperature consequently remains stationary.
Afterward, the core is essentially isothermal and heat from the curst is now rapidly flowing into the cold core,
a process which takes little more than 100 years and is reflected by a sudden drop of $T_e$
when the cooling wave reaches the surface.
After this, during phases C and D, the stellar interior is isothermal and it is only within the shallow envelope,
not shown in the Figure, that a temperature gradient is still present.

Notice that the models with pairing have a shorter crust relaxation time: this is simply due to the strong reduction of
the neutron specific heat in the inner crust by the $^1\! S_0$ gap.
\\
{\bf Phase C:} during this phase, the ``neutrino cooling phase'', the star evolution is driven by neutrino emission
from the core since $L_\nu \gg L_\gamma$.
The difference between ``slow'' and ``fast'' neutrino emission is now clearly seen.

Very noticeable is the effect of the pairing suppression of the neutrino emissivity in the fast cooling models:
once $T$ dropped below $T_c$, which happened only a few seconds/minutes after the beginning of the simulation,
neutrino cooling was quenched. 
Notice that the difference between the various $\epsilon_{Fast}^{(q)}$ processes is much smaller in the 
presence of pairing than in its absence:
that it took half a minute (for $q=27$) or half an hour (for $q=25$) for $T$ in the ``pit'' to drop below $T_c$
does not make much of a difference when looking at the star thousands of years later.
The evolution is more dependent on $T_c$ than on the actual value of $q$.

Considering the slow cooling models, comparison of the normal case with the paired one, but with the
PBF processed turned off, shows the same effect of pairing,
but less dramatic since the star is not as cold as in the fast cooling models.
However, when the PBF process is not artificially turned off the paired model is very similar to the normal one:
the burst of neutrino emission occurring when $T \simeq T_c$ from the constant formation and breaking of
Cooper pairs induces some extra, temporary, cooling.
The impact of the PBF process, however, depends on the size of the neutron $^3\! P\! - \! F_2$ gap
and is considered in more detail in the next section.
\\
{\bf Phase D:} at late times, $L_\nu$ has dropped significantly due to its strong $T$ dependence and
photon emission, $L_\gamma$, now drives the evolution. 
This is seen in the cooling curves by their very large slopes.
During this ``photon cooling era'' the models with pairing cool faster due to the reduction of the specific heat.

\section{The Minimal Cooling Paradigm and Cas A}

Table~\ref{Tab:nu} shows that there are many possibilities for having some fast neutrino emission and, from the results 
of the previous section, this implies that any observed $T_e$ could be fit {\em if} there is a gap of the appropriate size that
controls the neutrino emission.
However, a simple question arise: is there any observational evidence for the occurrence of fast neutrino cooling?
If yes, how strong is it?
To address this question, the ``Minimal Cooling'' paradigm was developed \cite{Page:2004zr,Page:2009vn}
which is a natural extension of the previous ``Standard Cooling'' (see, e.g., \cite{Nomoto:1987kx}).
The essence of the minimal cooling paradigm is the {\em a priori} exclusion of all fast neutrino emission processes.
Core neutrino cooling is, hence, limited to the modified Urca with the similar nucleon bremsstrahlung processes,
and the PBF process.
Minimal cooling is, however, not naive cooling:
it takes into account all other uncertainties on the microphysics and the astrophysical conditions.
The major factors turn out to be the size of the neutron triplet-pairing gap and the chemical composition of the envelope.

\begin{figure}[b]
\centerline{\psfig{file=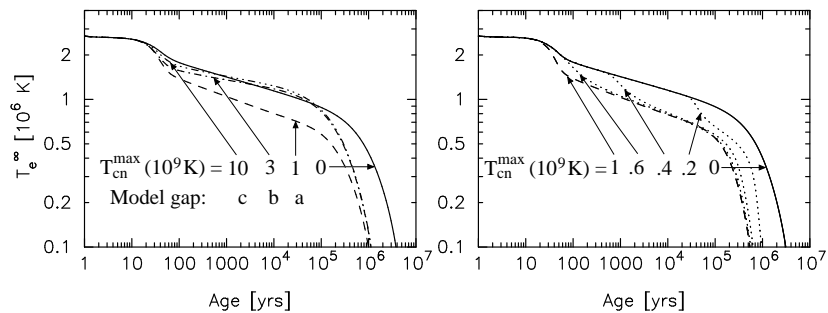,width=0.95\textwidth}}
\caption{Gauging the effect of neutrino emission by Cooper pair formation (the PBF process) within minimal cooling.
               Large values, $\geq 10^9$ K, of $T_{c \, n}^\mathrm{max}$ for the neutron $^3\! P\! - \! F_2$ gap 
               are considered in the left panel
               while smaller values, $\leq 10^9$ K, are considered in the right panel.
               See text for description.}
\label{Fig:Vary_3P2}
\end{figure}

It was shown in Figure~\ref{Fig:Cool_Basic} that, with a relatively large value of the neutron $^3\! P\! - \! F_2$ gap,
the PBF process can compensate the suppression of the modified Urca process. 
A further study of this effect is presented in Figure~\ref{Fig:Vary_3P2}.
In the left panel, a cooling curve with a vanishing $^3\! P\! - \! F_2$ gap and three curves with the gaps ``a'', ``b'', and ``c''
of Figure~\ref{Fig:Tc_n} are presented.
The all-important parameter is the maximum value of $T_c$ reached in the core, $T_{c n}^\mathrm{max}$, 
since the pairing phase transition will start when the core temperature $T$ reaches $T_{c n}^\mathrm{max}$.
It is seen that colder stars during the neutrino cooling era are obtained with a value $T_{c n}^\mathrm{max} = 10^9$~K
than for larger values.
The reason for this result is that with such a gap as model ``a'' there is always a significant region of the core that is going through
the phase transition at ages between $10$ to $10^5$ years while in the cases of the larger gaps ``b'' or ``c'' the pairing phase transition
occurred within most of the core at much earlier times and most of the neutrino emission, including the one from the PBF process,
is hence suppressed at later ages, resulting in warmer stars during the neutrino cooling era.
The right panel of Figure~\ref{Fig:Vary_3P2} describes the effect of smaller neutron $^3\! P\! - \! F_2$ gaps:
they are the model gap ``a'' scaled by a factor $x = 0.6$, $0.4$, or $0.2$, resulting in $T_{c n}^\mathrm{max} = x \times10^9$ K.
Ones sees that the cooling trajectories for $x <1$ separate from the upper trajectory, with no $^3\! P\! - \! F_2$ gap, 
at a point which is precisely the moment when the core temperature $T$ reaches $T_{c n}^\mathrm{max}$ 
and the pairing phase transition
starts \footnote{A model with $x=0.8$ would be essentially undistinguishable from the case $x=1$ as the phase transition would
start before the end of the crust relaxation phase.}.
This late onset of the pairing phase transition results in a transitory period of accelerated cooling.

\subsection{Minimal cooling vs data}

\begin{figure}[t]
\centerline{\psfig{file=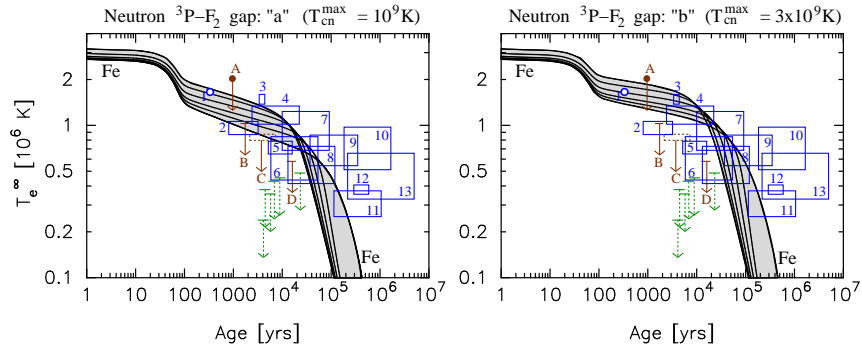,width=0.99\textwidth}}
\caption{Comparison of two minimal cooling scenarios with observational data.
               The neutron $^3\! P\! - \! F_2$ gaps employed are shown in Figure \ref{Fig:Tc_n}.
               See text for description.}
\label{Fig:Cool_Minimal_a-b}
\end{figure}

The present set of observational data on isolated cooling neutron stars is displayed in the 
Figures~\ref{Fig:Cool_Minimal_a-b} and \ref{Fig:Cool_Minimal_CasA}.
The first subset of stars, displayed as boxes and numbered 1 to 13, are objects from which a thermal spectrum,
in the soft X-ray band $E_X \sim 0.1 - 5$ keV, is clearly detected. 
For each case the plotted $T_e$ is obtained from the thermal luminosity using Eq.~(\ref{Eq:Lphot}).
The stars are:
1) CXO J232327.8+584842 (Cas A),
2) PSR J1119-6127 (G292.2-0.5),
3) PSR J0821-4300 (Puppis A),
4) PSR 1E1207.4-5209 (PKS 1209-52),
5) PSR B0833-45 (Vela),
6) PSR B1706-44 (G343.1-2.3),
7) PSR B0538+2817 (S147),
8) PSR B2334+61(G114.3+0.3),
9) PSR B0656+14,
10) PSR B1055-52,
11) PSR B0633+1748 aka ``Geminga'',
12) RX J1856.5-3754, and
13) RX J0720.4-3125
(in parenthesis is given the name of the associated supernova remnant).
The second subset consists of four pulsars whose soft X-ray spectrum is dominated by magnetospheric emission,
appearing as a power-law spectrum, so that only an upper limit on $T_e$ can be inferred.
They are:
A) PSR B0531+21 (Crab),
B) PSR J1124-5916 (G292.0+1.8),
C) PSR J0205+6449 (3C58), and
D) PSR J0007.0+7303 (CTA1).
Finally, six upper limits, shown as dotted arrows, are obtained from the total absence of detection of any object
in six supernova remnants:
some of these remnants may contain an isolated black hole, but it is unlikely to be the case for the six of them.
I refer the reader to \cite{Page:2004zr} and \cite{Page:2009vn} for details and discussion of these data points.

In Figure~\ref{Fig:Cool_Minimal_a-b} these data are compared with two series of models with a moderate (left panel)
and a large (right panel) neutron $^3\! P\! - \! F_2$ gap. 
The grey shaded areas, encompassing several cooling curves, show the range of predictions resulting from the
uncertainty on the chemical composition of the envelope: the curve marked ``Fe'' assumes the presence of iron peak nuclei 
while the other curves show the effect of having increasing amounts of light elements in the envelope
(the interior physics being the same for all curves). 
Stars with ages less than $10^5$ years are the most critical to infer properties of dense matter since
they are in the neutrino cooling era.
In the case $T_{c n}^\mathrm{max} = 10^9$ K, the theoretical predictions are compatible with all observed values and
upper limits, with the only exception of the PSR in CTA1 (D) if one does not consider the six non-detection from
supernova remnants.
On the other hand, for larger $T_{c n}^\mathrm{max}$ theoretical predictions are incompatible with the majority of data points
during the neutrino cooling era.

Given the uncertainty on the theoretical predictions of the size of the neutron $^3\! P\! - \! F_2$ gap, these results
are not very encouraging:
a moderate gap implies that most observed young neutron stars show no evidence of fast neutrino emission while
a large gap results in the opposite conclusion that most observed young neutron stars show evidence of fast neutrino
emission.

The older stars, 10 to 13, clearly appear warmer that the theoretical predictions:
they likely imply the presence of some ``heating'' mechanism, the ``$H$'' term in Eq. (\ref{Eq:cooling})
(see, e.g., \cite{Page:2006fk}).

\subsection{The cooling of the neutron star in Cassiopeia A}

\begin{figure}[t]
\centerline{\psfig{file=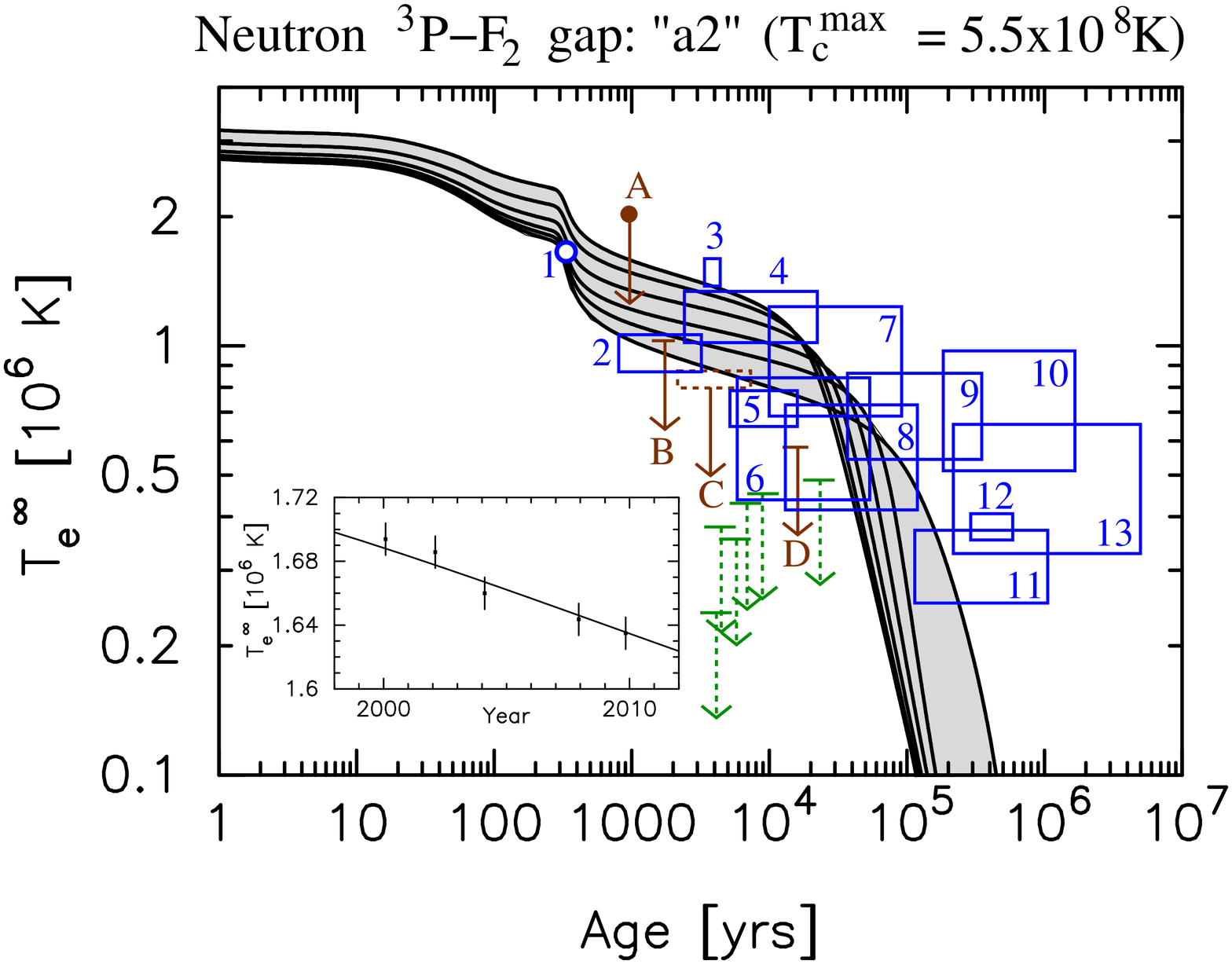,width=0.49\textwidth}
\psfig{file=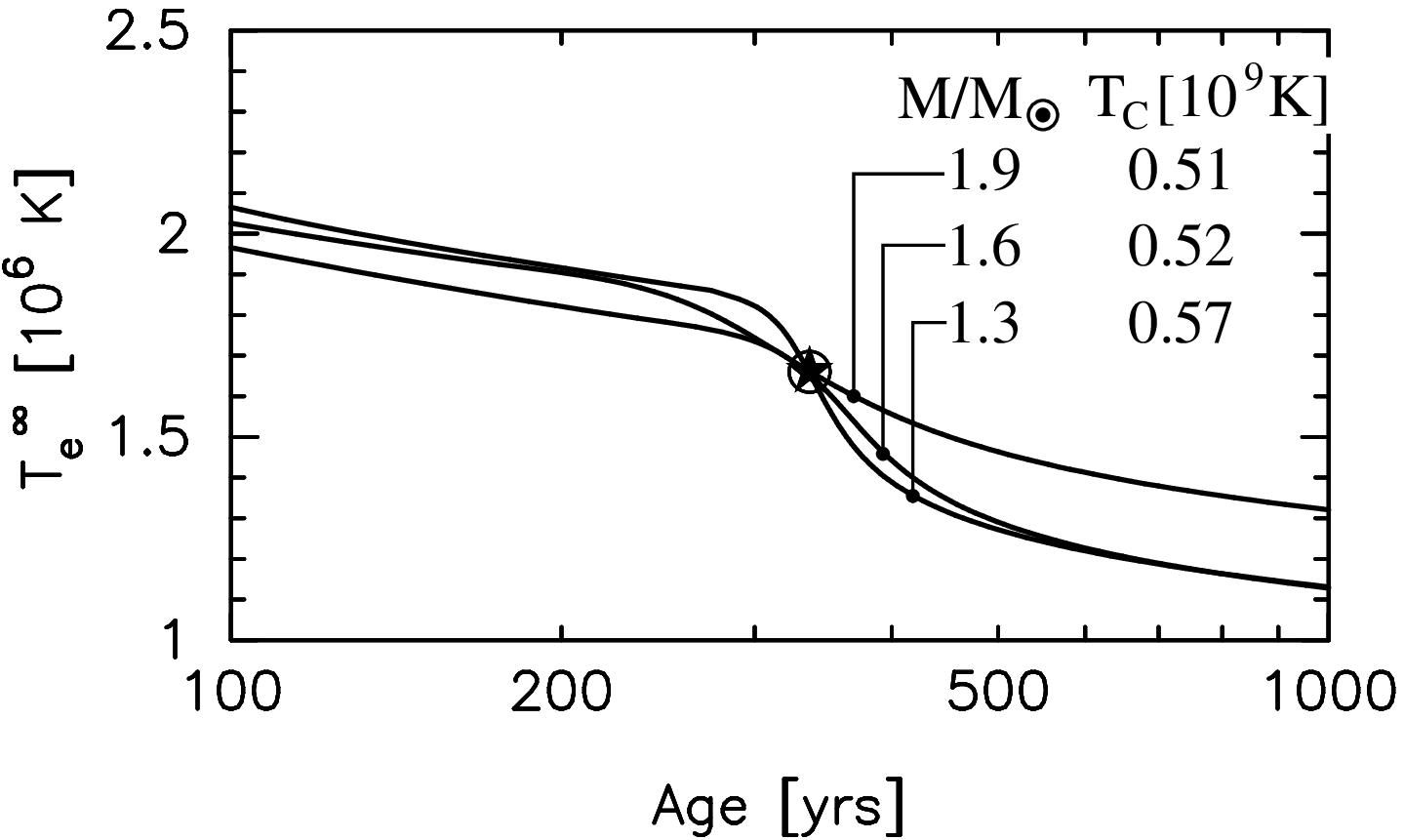,width=0.45\textwidth}}
\caption{Left panel: a minimal cooling scenario reproducing the observed rapid cooling of Cas A (data point 1).
               The inset, from \cite{Page:2011ys}, shows the fit to the five temperatures reported in \cite{Heinke:2010zr}.
               Right panel: dependence of the slope $s$ of the cooling curve on the star mass:
               $s = 0.5$, $0.9$, and $1.4$ for $M = 1.3$, $1.6$, and $1.9 M_\odot$, resp., at 330 yrs (from \cite{Page:2011ys}).}
\label{Fig:Cool_Minimal_CasA}
\end{figure}

To be able to distinguish between the various possible neutron $^3\! P\! - \! F_2$ gaps some new observational datum(a) is needed.
This has likely been recently provided by the observation of the cooling, in real time, of the neutron star in the
Cassiopeia A supernova remnant \cite{Heinke:2010zr}.
Between the year 2000 till 2009 this neutron star appears to have cooled by 4\% (with a 20\% decrease in flux).
The observed thermal luminosity of Cas~A is $L_\gamma \simeq 10^{34}$~erg~s$^{-1}$:
from the observed cooling rate, and a simple estimate of its specific heat, Cas~A must be loosing about
$10^{38}$~erg~s$^{-1}$, in the case the cooling represents the evolution of an isothermal star.
It is hard to imagine that such an enormous cooling rate is a persistent evolution, and it is more likely that
``something critical'' happened recently:
for a cooling neutron star this ``something'' should be the temperature and a critical temperature means a phase transition.
The right panel of Figure~\ref{Fig:Vary_3P2} precisely exhibits such sudden increase in the cooling rate when
the star's core temperature reaches $T_{c n}^\mathrm{max}$.
With the known age of Cas A of 330 years
\footnote{The supernova explosion was likely observed by J. Flamsteed on 16 August 1680.},
the right panel of Figure~\ref{Fig:Vary_3P2} indicates a $T_{c n}^\mathrm{max}$ of the order
of $0.5 \times 10^9$ K.
This interpretation was proposed in \cite{Page:2011ys} and, simultaneously and independently, in \cite{Shternin:2011ly}.
The left panel of Figure~\ref{Fig:Cool_Minimal_CasA} is an  up-date of Figure~\ref{Fig:Cool_Minimal_a-b}
using the neutron $^3\! P\! - \! F_2$ gap model ``a2'' of Figure~\ref{Fig:Tc_n} as proposed in \cite{Page:2011ys}.

The deduced value of $T_{c n}^\mathrm{max}$ is mostly fixed just by the known age of the star.
Reproducing the observed cooling slope, $s=-d\log T_e/d \log t\simeq 1.2$, however, requires the star was quite hot
before the onset of the phase transition, which means
its previous neutrino luminosity was low. 
Suppression of $L_\nu$ is naturally obtained if protons were already superconducting, with a higher $T_c$.
Obtaining $s \simeq 1.2$ requires proton superconductivity in a significant part of the core, and this observed slope
would also put strong constraints on proton superconductivity:
results of Figure~\ref{Fig:Cool_Minimal_CasA} assume the proton $^1\! S_0$ gap ``CCDK'' of Figure~\ref{Fig:Tc_p}.
Within the set of proton $^1\! S_0$ gaps presented in Figure~\ref{Fig:Tc_p} this model is the one reaching to the
highest densities.
The first report of \cite{Heinke:2010zr} was subsequently consolidated by another observation reported in
\cite{Shternin:2011ly}.
However, there are delicate calibration issues \cite{Elshamouty:2012fk} and the actual slope may be smaller,
a possibility that would ease the strong requirement on the proton $^1\! S_0$ gap and/or allow for a more massive star.
The right panel of Figure~\ref{Fig:Cool_Minimal_CasA} shows the change in the evolution when increasing the
neutron star mass:
a larger mass simply results in having a larger portion of the core where protons are not superconducting
and, hence, a higher initial $L_\nu$ and a smaller slope during the rapid cooling phase.
Similar results are obtained by changing the density range covered by proton superconductivity while keeping
the star mass constant.

\section{Conclusions}

If the cooling of Cas A is confirmed by future observations, it is certainly one of the most amazing pieces of observational
data on neutron stars and the interpretation presented here would mean that we are seeing, in real time, about a solar
mass of neutrons going through the triplet pairing phase transition.
It would also imply a neutron $^3\! P\! - \! F_2$ energy gap of the order of 0.1 MeV.

In a more general context, almost any cooling scenario compatible with observational data needs some type of pairing,
either for protons, neutrons or exotica.
A possible exception being the ``medium-modified Urca'' scenario \cite{Voskresensky:2001ys}.
Given the large number of possible scenarios it is difficult, when going beyond the minimal cooling paradigm, to
decide which type of exotica is acting (see, e.g., \cite{Page:2000uq}).
However, some cold young neutron stars, as the one in CTA1, tell us that more than minimal cooling is definitely needed.

\smallskip
{\bf Acknowledgments:} this work was supported by grants from 
Conacyt (CB-2009/132400) and UNAM-DGAPA (PAPIIT, \# IN 113211).
The author is grateful to his collaborators, M. Prakash, J. M. Lattimer and A. S. Steiner, 
but the present text is under his own responsibility.

\bibliographystyle{ws-rv-van} 


\printindex                         

\end{document}